\documentclass{czjphyse}       
\usepackage{epsf}  
\begin{document}
\title{SPIN EFFECTS IN DIFFRACTIVE REACTIONS}        
\authori{S.V.Goloskokov}      
\addressi{Bogoliubov Laboratory of Theoretical  Physics,
 Joint Institute for Nuclear Research,
Dubna 141980, Moscow region, Russia}     
\authorii{}     
\addressii{}    
\authoriii{}    
\addressiii{}   
\headtitle{ Spin Effects In Diffractive Reactions \ldots}            
\headauthor{S.V.Goloskokov}           
\specialhead{S.V.Goloskokov:Spin Effects In Diffractive Reactions  \ldots} 
\evidence{A}
\daterec{XXX}    
\cislo{0}  \year{2000} \setcounter{page}{1} \pagesfromto{000--000}
\maketitle

\begin{abstract}
Double spin asymmetries in the diffractive $Q \bar Q$ and
$J/\Psi$ leptoproduction are discussed. It is shown that the
asymmetries for longitudinally polarized lepton and
longitudinally or transversely  polarized proton can be used to
study spin dependent gluon distribution of the proton at small
$x$.
\end{abstract}

\section{Introduction}
In this report, we study diffractive processes at high energies
which are predominated by the Pomeron exchange. The Pomeron is a
color singlet object which can be associated with the two-gluon
state \cite{low}. The cross section of inclusive hadron
production is expressed in terms of ordinary parton distributions
where partons have the same momenta. The diffractive reactions,
which should give significant contribution to inclusive processes
at small Bjorken $x \leq 0.1$, have a different nature. Really, in
the diffractive hadron leptoproduction, like vector meson and $Q
\bar Q$ production, the nonzero momentum $x_P$ carried by the
two-gluon system (Pomeron) appears and the gluon momenta cannot
be equal. As a result, these processes can be expressed in terms
of skewed parton distribution (SPD) in the nucleon ${\cal
F}_\zeta(x)$ where $x$ is a fraction of the proton momentum
carried by the outgoing gluon and $\zeta$ is the difference
between the gluon momenta (skewedness) \cite{rad-j}.
Investigation of the diffractive reactions should play a key role
in future study  ${\cal F}_x(x)$ at small $x$. For the processes
which involve the light quark production both quark and gluon SPD
will contribute especially for not small $x$. We shall discuss
here the diffractive charm quark production and $J/\Psi$
production where the $q \bar q$ exchange in the $t$ -channel is
not important because the charm quark content in  the proton is
rather small. In these reactions the predominated contribution is
determined by the two-gluon exchange (gluon SPD). Analysis of
such diffractive reactions should throw light on the gluon
structure of the proton at small $x$ \cite{rysk,brod}.

To study spin effects in diffractive hadron production, one must
know the structure of the two-gluon coupling with the proton at
small $x$. In the leading Log approximation, the ladder gluon
graphs give predominating contribution. In a QCD--inspired diquark
model of the proton \cite{kroll}, such graphs have been analyzed
in \cite{gol_kr}. The model generates the spin-dependent $ggp$
coupling at moderate momentum transfer. Using the results of
\cite{gol_kr}, the following form of the coupling has been found
\begin{eqnarray}\label{ver}
V_{pgg}^{\alpha\alpha'}(p,t,x_P,l_\perp)&=& (\gamma^{\alpha}
p^{\alpha'} + \gamma^{\alpha'} p^{\alpha}) B(t,x_P,l_\perp)+4
p^{\alpha} p^{\alpha'}
A(t,x_P,l_\perp)\nonumber\\
&+&\epsilon^{\alpha\beta\delta\rho}p_{\delta}\gamma_{\rho}\gamma_{5}
D(t,x_P,l_\perp).
\end{eqnarray}
Here the functions $A-D$ have dependence on the transverse part of
the gluon momentum $l_\perp $.

The structure $(\gamma^{\alpha} p^{\alpha'} + \gamma^{\alpha'}
p^{\alpha}) B(t)$ in (\ref{ver}) is a coupling which determines
the spin-non-flip contribution. The term $p_{\alpha}p_{\alpha'}
A(r)$ leads to the transverse spin-flip in the  vertex which does
not vanish in the $s \to \infty$ limit. The first two terms of the
vertex (\ref{ver}) are symmetric over the $\alpha,\alpha'$
indices. They are equivalent to the isoscalar electromagnetic
nucleon current with the Dirac and Pauli nucleon form factors
\cite{nach}.  In the model \cite{gol_mod}, the form (\ref{ver})
was found to be valid for the momentum transfer $|t| < \mbox{few
GeV}^2$ and the $A(t)$ amplitude was caused  by the meson-cloud
effects in the nucleon. Within this model, a quantitative
description of meson-nucleon and nucleon--nucleon polarized
scattering at high energies has been obtained \cite{gol_mod}. The
model predictions for polarization for the PP2PP experiment at
RHIC was made in \cite{akch}. The expected errors are quite small
and the information about the spin-flip part of the coupling
(\ref{ver}) can be obtained experimentally. In a QCD--based
diquark model  the $A(t)$ contribution is determined by the
effects of vector diquarks inside the proton \cite{gol_kr}, which
are of an order of $\alpha_s$. Diquarks provide an effective
description of nonperturbative effects in the proton. The single
spin transverse asymmetry predicted in the models
\cite{gol_kr,gol_mod,akch} is about 10\% for $|t| \sim 3
\mbox{GeV}^2$ which is of the same order of magnitude as has been
observed experimentally \cite{krish-f}. These model approaches are
consistent with experiment and give the ratio
\begin{equation}\label{r}
\frac{|A(t,x_P)|}{|B(t,x_P)|}\leq 0.1 \mbox{GeV}^{-1}.
\end{equation}
The asymmetric structure in (\ref{ver}) is proportional to $D
\gamma_{\rho}\gamma_{5}$ and can be associated with $\Delta G$.
It should give a visible contribution to the  double spin
longitudinal asymmetry $A_{ll}$ \cite{bartels}. Unfortunately, it
does not appear for elastic scattering ($x_P=0$) and the value of
this structure is not well known from our model estimations.

In this report, we shall analyze spin effects caused by the
structures $A$ and $B$. It will be shown that such effects will
be small in the $A_{ll}$ asymmetry. This means that, most likely,
they  do not provide additional problems with extracting the
$\Delta G \propto D \gamma_{\rho}\gamma_{5} $ term from the
$A_{ll}$ asymmetry in the hadron leptoproduction at small $x$. It
will be shown that the double spin asymmetry for longitudinally
polarized lepton and transversely polarized proton is mainly
determined by the $A$ term in (\ref{ver}). This asymmetry should
be used to study this structure in the $ggp$ coupling.

\section{Vector meson production and skewed parton distribution}
Let us study the diffractive $J/\Psi$ production at high energies
$s=(p+l_e)^2$ and fixed momentum transfer $t= r_P^2=-\vec
\Delta^2=(p-p')^2$. Here $l_e$ and $p$ are the initial momenta of
the lepton and proton, $p'$ is the final proton momentum and
$r_P$ is a momentum carried by the Pomeron ($\vec \Delta$ is its
transverse part). The vector meson production amplitude is
described in addition to $s$ and $t$ by the variables
\begin{equation}
\label{momen}
\nonumber \\
Q^2=-q^2,\quad
 y=\frac{p \cdot q}{p \cdot l_e},\quad x_P=\frac{q\cdot r_P}{q \cdot
p},
\end{equation}
where $Q^2$ is the photon virtuality. The variables $y$ and $x_P$
determine the fractions of the longitudinal momenta of the lepton
and proton carried by the photon and Pomeron, respectively. From
the mass-shell equation for the vector--meson momentum
$V^2=(q+r_P)^2=M_J^2$ we find that in this reaction
\begin{equation}
\label{x_P} x_P \sim \frac{m_J^2+Q^2+|t|}{s y}
\end{equation}
and it is small at high energies.

A simple model is considered for the amplitude of the
$\gamma^\star \to J/\Psi$ transition.  The virtual photon is
going to the $q \bar q$ state and the $q \bar q \to V$ amplitude
is described by a non-relativistic wave function
\cite{rysk,diehl}. In this approximation, quarks have the same
momenta $k$  equal to half of the vector meson momentum and mass
$m_c=m_J/2$. Gluons from the Pomeron are coupled with the single
and different quarks in the $c \bar c$ loop. This ensures the
gauge invariance of the final result.

The spin-average and spin dependent cross sections of the
$J/\Psi$ leptoproduction with parallel and antiparallel
longitudinal polarization of a lepton and a proton are determined
by the relation
\begin{equation}\label{ds0}
d \sigma(\pm) =\frac{1}{2} \left( d \sigma(^{\rightarrow}
_{\Leftarrow}) \pm  d \sigma(^{\rightarrow}
_{\Rightarrow})\right).
\end{equation}
The cross section $d \sigma(\pm)$ can be written in the form
\begin{equation}\label{ds}
\frac{d\sigma^{\pm}}{dQ^2 dy dt}=\frac{|T^{\pm}|^2}{32 (2\pi)^3
 Q^2 s^2 y}.
\end{equation}
 For the spin-average  amplitude square we find \cite{golj}
\begin{equation}\label{t+}
|T^{+}|^2=  s^2\, N \,\left( (2-2 y+y^2) m_J^2 + 2(1 -y) Q^2
\right) \left[ |\tilde B+2 m \tilde A|^2+|\tilde A|^2 |t| \right].
\end{equation}
 Here the term proportional to $(2-2 y+y^2) m_J^2$ represents
the contribution of the virtual photon with transverse
polarization. The $2(1 -y) Q^2$ term describes the effect of
longitudinal photons. The $N$ factor in (\ref{t+}) is
normalization, and the $\tilde A$ and $\tilde B$ functions are
expressed through the integral over transverse momentum of the
gluon. The function $\tilde B$ is determined by
\begin{eqnarray}\label{bt}
\tilde B&=&\frac{1}{4 \bar Q^2} \int \frac{d^2l_\perp
(l_\perp^2+\vec l_\perp \vec \Delta) B(l_\perp^2,x_P,...)}
{(l_\perp^2+\lambda^2)((\vec l_\perp+\vec
\Delta)^2+\lambda^2)[l_\perp^2+\vec l_\perp \vec \Delta
+\bar Q^2]} \nonumber\\
&\sim& \frac{1}{4 \bar Q^4}\int^{l_\perp^2<\bar Q^2}_0
\frac{d^2l_\perp (l_\perp^2+\vec l_\perp \vec \Delta) }
{(l_\perp^2+\lambda^2)((\vec l_\perp+\vec \Delta)^2+\lambda^2)}
B(l_\perp^2,x_P,...).
\end{eqnarray}
with $\bar Q^2=(m_J^2+Q^2+|t|)/4$. The term $(l_\perp^2+\vec
l_\perp \vec \Delta)$ appears in the numerator of (\ref{bt})
because of the cancellation between the graphs where gluons are
coupled with the single and different quarks. The $\tilde A$
function is determined by the similar integral.

The integral (\ref{bt}) is  found to be connected with the gluon
SPD
\begin{equation}\label{fspd}
{ {\cal F}^g_{x_P}(x_P,t)}=\int^{l_\perp^2<\bar Q^2}_0
\frac{d^2l_\perp (l_\perp^2+\vec l_\perp \vec \Delta) }
{(l_\perp^2+\lambda^2)((\vec l_\perp+\vec \Delta)^2+\lambda^2)}
B(l_\perp^2,x_P,...).
\end{equation}
We find that $B(l_\perp^2,x_P,...)$ is the unintegrated spin-
average gluon distribution. The $\tilde A$ function can be
determined in terms of the ${\cal K}^g_{x_P}(x_P,t)$ distribution.
Determination of the gluon distribution functions can be found in
\cite{kroll-da}, e.g.

The spin-dependent amplitude square looks like
\begin{equation}\label{t-}
|T^{-}|^2= s |t|  (2- y) N   \left[ |\tilde B|^2+ m (\tilde
A^\star \tilde B +\tilde A \tilde B^\star) \right] m_J^2.
\end{equation}
\begin{figure}
\centering \mbox{\epsfxsize=90mm\epsffile{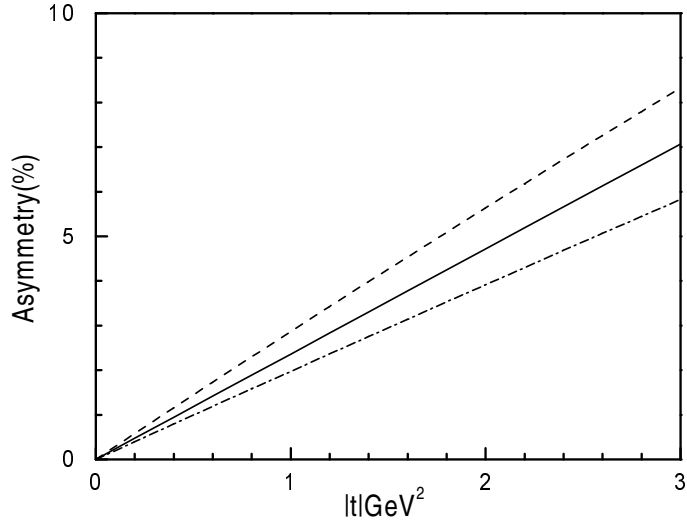}}
\caption{The $A_{ll}$ asymmetry of the $J/\Psi$ production  at
HERMES: solid line -for $\alpha_{flip}=0$; dot-dashed line -for
$\alpha_{flip}=-0.1$; dashed line -for $\alpha_{flip}=0.1$}
\label{F1}
\end{figure}

The $A_{ll}$ asymmetry for not high $Q^2$ is determined by
\begin{equation}
\label{asy} A_{ll}=\frac{\sigma(-)}{\sigma(+)} \sim
\frac{|t|}{s}\frac{(2-y) \left[ |\tilde B(t)|^2+ m (\tilde
A(t)^\star \tilde B(t)+ \tilde A(t) \tilde B(t)^\star) \right]}
{(2-2 y+ y^2)\left[ |\tilde B(t)+2 m \tilde A(t)|^2+ |t| |\tilde
A(t)|^2 \right]}.
\end{equation}
 It has been shown in \cite{gola_ll} that the
$A_{ll}$ asymmetry in the diffractive processes is proportional
to $x_P$. For the diffractive  vector meson production $x_P \sim
1/s$ see (\ref{x_P}). As a result, the obtained $A_{ll}$ asymmetry
decreases with growing energy.

The $A_{ll}$ asymmetry also depends on the ratio of the spin-flip
to the non-flip  parts of the  coupling (\ref{ver})
$\alpha_{flip}=\tilde A(t)/ \tilde B(t)$ which has been found in
(\ref{r}) to be about 0.1. The predicted asymmetry at HERMES
energies is shown in Fig.\ 1. The contribution of the
spin-dependent $A$ term in (\ref{ver}) to the double spin
$A_{ll}$ asymmetry of the $J/\Psi$ production does not exceed two
per cent for the momentum transfer $|t| \leq 1 \mbox{GeV}^2$.
Sensitivity of the asymmetry to $\alpha$ is rather weak.   At
HERA energies, the asymmetry will be negligible.

\section{Spin effects in $Q \bar Q$ production}
Now we are interested in the diffractive $Q \bar Q$
leptoproduction. This reaction might be determined in terms of
variables (\ref{momen}). However, the mass of the produced hadron
system is not fixed here and an additional variable $\beta= x/x_P$
appears. In the two-gluon picture of the Pomeron we consider the
graphs of Fig. 2.
\begin{figure}
\centering \mbox{\epsfysize=20mm\epsffile{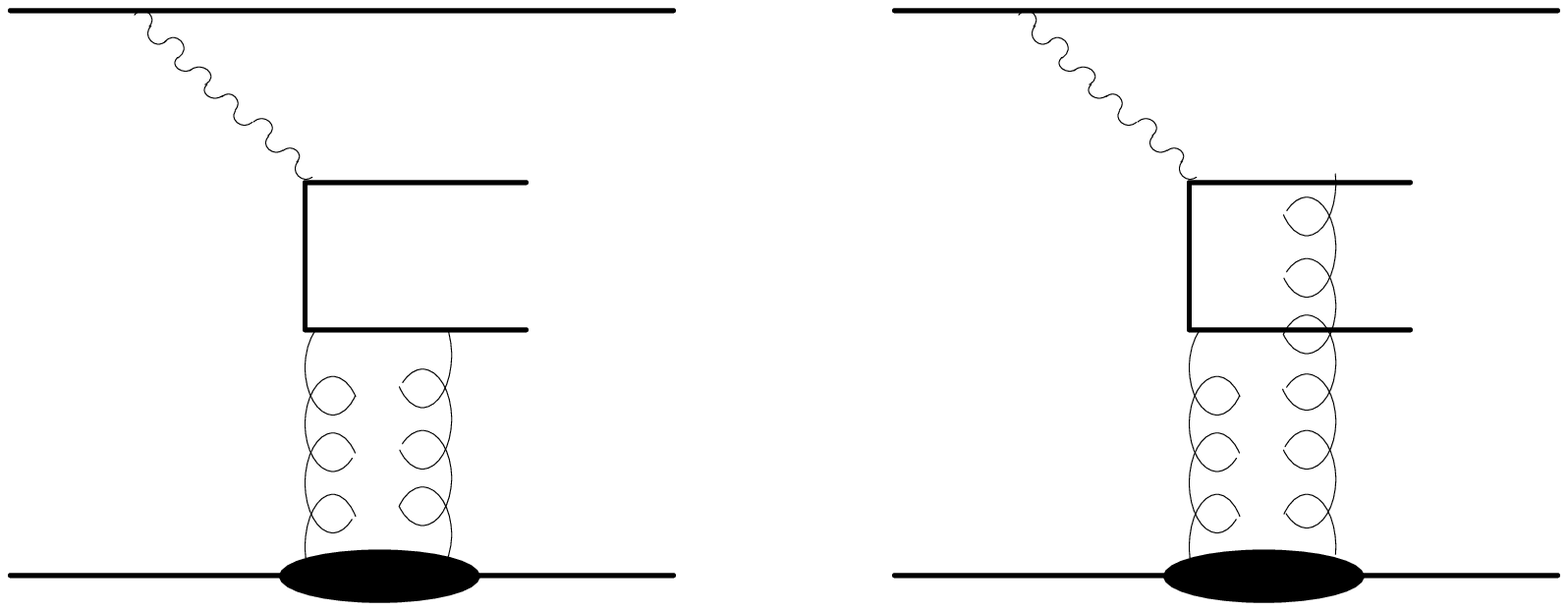}}
\caption{Two-gluon contribution to $Q \bar Q$ production}
\label{F2}
\end{figure}
The cross section is determined by the square of the graphs shown
in Fig. 2, and includes, together with the planar, the nonplanar
contribution with the crossed quark lines. The spin-average and
spin-dependent cross section can be written in the form
\begin{equation}
\label{sigma} \frac{d^5 \sigma(\pm)}{dQ^2 dy dx_P dt dk_\perp^2}=
\left(^{(2-2 y+y^2)} _{\hspace{3mm}(2-y)}\right)
 \frac{C(x_P,Q^2) \; N(\pm)}
{\sqrt{1-4k_\perp^2\beta/Q^2}}.
\end{equation}
Here $C(x_P,Q^2)$ is a normalization function which is common for
the spin average and spin dependent cross section. The planar
contribution to $N^{+}$ (sum of diagrams with non-crossed quark
lines) looks like
\begin{equation}\label{N1}
N^{+}_P=(m^2_c+(\vec k_\perp+\vec r_\perp)^2) \left( |\tilde B+2 m
\tilde A|^2+|\tilde A|^2 |t| \right)
\end{equation}
with
\begin{eqnarray}\label{btn}
\tilde B&=&\frac{1}{\bar Q_1^2} \int \frac{d^2l_\perp
(l_\perp^2+\vec l_\perp \vec \Delta-\vec l_\perp \vec k_\perp)
B(l_\perp^2,x_P,...)} {(l_\perp^2+\lambda^2)((\vec l_\perp+\vec
\Delta)^2+\lambda^2)[(\vec l_\perp-\vec k_\perp)^2+m_c^2]} \nonumber\\
&\sim& \frac{1}{\bar Q_1^4}\int^{l_\perp^2<\bar Q^2_1}_0
\frac{d^2l_\perp (l_\perp^2+\vec l_\perp \vec \Delta) }
{(l_\perp^2+\lambda^2)((\vec l_\perp+\vec \Delta)^2+\lambda^2)}
B(l_\perp^2,x_P,...) \nonumber\\
&=&\frac{1}{\bar Q_1^4} {\cal F}^g_{x_P}(x_P,t,\bar Q^2_1)
\end{eqnarray}
where $Q^2_1=m_c^2+\vec k_\perp^2$ and $k$ is a quark momentum.

The calculation shows considerable cancellation between the
contributions where gluons are coupled with the single and
different quarks (Fig. 2). As a result,  the numerator of the
integral for the function $\tilde B$ (\ref{btn}) is vanishing
when the gluon momentum $l_{\perp}$ is going to zero as in the
vector meson production. Corresponding integrals for $\tilde A$
have a similar form. We see that the cross section of the
diffractive $Q \bar Q$ production is expressed through the same
skewed gluon distributions as the vector meson production but at
a different scale.
\begin{figure}
\centering \mbox{\epsfxsize=90mm\epsffile{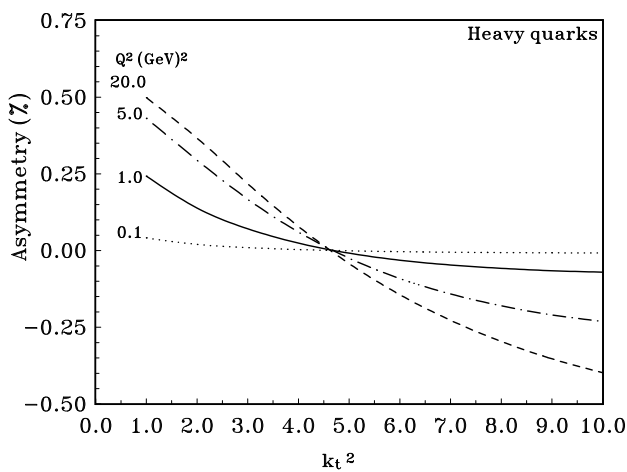}}
\caption{The predicted $Q^2$ dependence of the $A_{ll}$ asymmetry
for the $c \bar c$ production at COMPASS for $\alpha=0.1
\mbox{GeV}^{-1}$, $x_P$=0.1, $y$=0.5} \label{F3}
\end{figure}

Similar calculations have been made for the spin-dependent part of
the cross section $\sigma(-)$. We shall discuss here the
$A_{ll}=\sigma(-)/\sigma(+)$ asymmetry at COMPASS \cite{compass}.
 The obtained asymmetry is proportional to
$x_P$ (here $x_P$ is typically of about $.05-.1$)  and has a weak
energy dependence, Fig. 3. The predicted asymmetry is quite small
and does not exceed 1\%. It has a week dependence on the
$\alpha=\tilde A/\tilde B$ ratio. The $Q^2$ dependence of the
$A_{ll}$ asymmetry of the diffractive open charm production can
be estimated as $A_{ll} \propto Q^2/(Q^2+Q^2_0)$ with $Q^2_0 \sim
1 \mbox{GeV}^2$.
\begin{figure}
\centering \mbox{\epsfxsize=90mm\epsffile{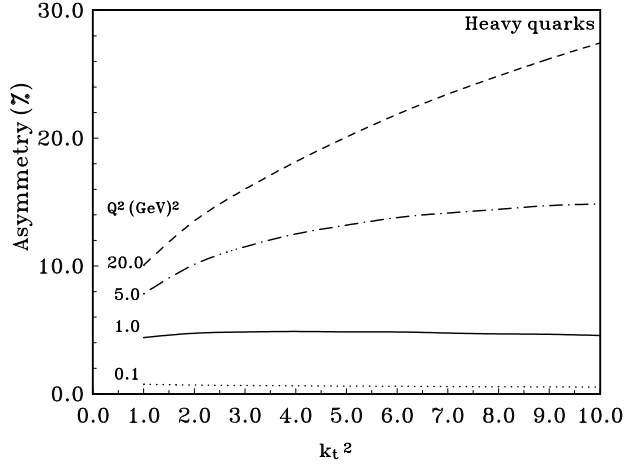}}
\caption{The predicted $Q^2$ dependence of the $A_{lT}$ asymmetry
for the $c \bar c$ production at COMPASS for
$\alpha=0.1\mbox{GeV}^{-1}$, $x_P$=0.1, $y$=0.5} \label{F4}
\end{figure}

We know from the study of elastic scattering that the $A$
structure in (\ref{ver}) is mainly responsible for the transverse
asymmetry. Not small single-spin asymmetry in $Q \bar Q$
production caused by the $A$ term has been predicted in
\cite{gol-ss}. This term should contribute to the $A_{lT}$
asymmetry with longitudinal lepton and transverse proton
polarization. The calculation of this asymmetry is similar to the
analysis of $A_{ll}$ which was made before. It has been found that
the $A_{lT}$ asymmetry is not small and proportional to the
$\alpha=\tilde A/\tilde B$ ratio. Also the $A_{lT}$ asymmetry is
proportional to the scalar production of the proton spin vector
and the jet momentum $A_{lT} \propto (s_{\perp} \cdot k_{\perp})
\propto \cos (\phi_{Jet})$. Thus, the asymmetry integrated over
the azimuthal jet angle $\phi_{Jet}$ is zero. We have calculated
the $A_{lT}$ asymmetry for the case when the proton spin vector
is perpendicular to the lepton scattering plane and the jet
momentum is parallel to this spin vector. The predicted asymmetry
is show in Fig. 4. It is huge and has a drastic $k^2_{\perp}$
dependence. The large value of the $A_{lT}$ asymmetry is caused
by the fact that in contrast to $A_{ll}$, it is not proportional
to  small $x_P$.
\section{Conclusion}
In this report the double spin asymmetries in the diffractive
hadron leptoproduction have been investigated using the two-gluon
picture of the Pomeron. We have considered all the graphs where
the gluons from the Pomeron couple to a different quark in the
loop and to the single one. This provides a gauge-invariant
scattering amplitude. Our calculations show that the
spin--dependent structure $A$ in (\ref{ver}) does not provide a
considerable contribution to the $A_{ll}$ asymmetry. The predicted
$A_{ll}$ asymmetry is smaller than 1-2\%. Not small effects in the
double spin $A_{ll}$ asymmetry should be determined by the
$\Delta G \propto D \gamma_{\rho}\gamma_{5}$ term of the vertex
(\ref{ver}). This contribution was not investigated here but can
be important for Compass experiment at small $x \leq 0.1$.
However, the results obtained here show that the diffractive
asymmetry in the $Q \bar Q$ production vanishes as $Q^2 \to 0$. We
hope that this conclusion is true for the $\Delta G$ term, too. If
so, such effects, most likely, do not provide additional troubles
in extracting $\Delta G$ from the $A_{ll}$ asymmetry. The reason
is that the COMPASS experiment is planing to study events of the
charm quark production at small $Q^2$ where the diffractive
asymmetry is expected be extremely small.

It is shown here that the $A_{lT}$ asymmetry of the diffractive
heavy quark production is determined by the $A$ structure in
(\ref{ver}). This asymmetry is predicted to be not small, about
10-20\%. It can give direct information about the spin-dependent
structure $A$ in the $ggp$ coupling. Note that it is usually
difficult to detect an outgoing proton in fixed target
experiments. Then, the cross section integrated over the momentum
transfer $t$ are measured. Our analyses show that such an
integrated asymmetry is smaller by a factor of 1.5--2 than the
unintegrated values shown in Figs 3, 4.

This report was supported in part by the Blokhintsev--Votruba
program and the Russian Fond of Fundamental Research, Grant
00-02-16696.

\end {document}